%% file: main.tex
\documentclass[letterpaper, 10 pt, conference]{ieeeconf}  

\usepackage{color}

\IEEEoverridecommandlockouts                              
\usepackage{algorithm}
\usepackage{algorithmic}
\usepackage{url}
\usepackage{graphicx}
\usepackage{subfigure}
\usepackage{amsmath,amssymb}

\title{\LARGE \bf
Social Bots for Online Public Health Interventions }
\author{\small Ashok Deb, Anuja Majmundar${^*}$, Sungyong Seo${^*}$, Akira Matsui,  Rajat Tandon, Shen Yan, Jon-Patrick Allem, and Emilio Ferrara
\thanks{*These authors contributed equally to this work.}
\\
University of Southern California, Los Angeles
\\
{Email: \{{adeb, anuja.majmundar, sungyons, amatsui,  rajattan, shenyan, allem, emiliofe\}}@usc.edu}
}

\begin{document}

\maketitle
\thispagestyle{empty}
\pagestyle{empty}

\begin{abstract}
According to the Center for Disease Control and Prevention, in the United States hundreds of thousands initiate smoking each year, and millions live with smoking-related diseases. Many tobacco users discuss their habits and preferences on social media. This work conceptualizes a framework for targeted health interventions to inform tobacco users about the consequences of tobacco use. We designed a Twitter bot named \textit{Notobot} (short for \textit{No-Tobacco Bot}) that leverages machine learning to identify users posting pro-tobacco tweets and select individualized interventions to address their interest in tobacco use. We searched the Twitter feed for tobacco-related keywords and phrases, and trained a convolutional neural network using over 4,000 tweets dichotomously manually labeled as either pro-tobacco or not pro-tobacco. This model achieves a 90\% recall rate on the training set and 74\% on test data. 
Users posting pro-tobacco tweets are matched with former smokers with similar interests who posted anti-tobacco tweets. Algorithmic matching, based on the power of peer influence, allows for the systematic delivery of personalized interventions based on real anti-tobacco tweets from former smokers. Experimental evaluation suggests that our system would perform well if deployed.  This research offers opportunities for public health researchers to increase health awareness at scale. Future work entails deploying the fully operational Notobot system in a controlled experiment within a public health campaign.
\end{abstract}
\input{src/intro}
\input{src/motivation}
\input{src/method}

\input{src/results}

\input{src/related}

\section{CONCLUSIONS}

We developed and presented a framework that leverages machine learning and artificial intelligence to assist public health researchers in conducting targeted and scalable interventions. Current practices either use population-level mass media interventions or develop interventions targeted mostly at specific demographic groups. A gap in public health intervention design efforts pertains to developing customized health awareness campaigns that identify health messages that are most likely to resonate with the target audience. Notobot, our notional design, leverages the persuasive power of social influence by identifying similar users in the Twitter network who post anti-tobacco tweets. Such an implementation can be scaled up at relatively low costs compared to traditional media campaigns. Notobot is based on a neural network for pro-tobacco use tweet identification and a decision tree for best-match intervention selection. The results presented here focus on the performance of the machine learning algorithms and the utility of Notobot as a concept design. Future work includes implementing and deploying Notobot in a public health campaign to test if the interventions selected perform better than common practice baseline.

\section*{APPENDIX}
\input{src/keywords}

\end{document}

%% file: src/intro.tex
\section{Introduction}
Cigarette smoking is the leading preventable cause of death in the United States~\cite{c1}\cite{c2}. This excess mortality among smokers is due to diseases related to smoking behaviors such as cancer and respiratory and vascular diseases~\cite{c1}. With the advent of social media, user exposure to and engagement with tobacco-related information is easier and faster than before. Social media users encounter pro-tobacco information more than anti-tobacco information on social media~\cite{c3}. Themes of pro-tobacco information on social media include images and/or videos promoting these products using cartoons~\cite{c45}, pairing hookah with alcohol in social settings~\cite{c41}, pairing of little cigar/cigarillos with marijuana~\cite{c42}, and showing smokers blowing large clouds of e-cigarette aerosol~\cite{c43}. Mere exposure to such tobacco-related content is known to be higher among youth susceptible to combustible tobacco use~\cite{c8}. Excess exposure to pro-tobacco social media messages can create low harm perceptions about tobacco use and/or normalize social norms of tobacco use in the society~\cite{c11}. The interactive nature of social media also allows its users, including adolescents, to access, create and share social media content related to tobacco use~\cite{c9}. Additionally, social media platforms can provide users with individualized experiences that give a more meaningful effect. Recent evidence suggests that, compared to adolescents who did not engage with tobacco-related content, those who engaged with such content reported higher incidences of tobacco initiation, increased frequency of use, and lower incidence of tobacco cessation at follow-up~\cite{c11}.    
Targeted smoking prevention communication strategies can address such imbalance in the nature of tobacco information (pro- vs. anti-tobacco). Compelling health messaging strategies of successful campaigns such as the Food and Drug Administration's (FDA) ``The Real Cost'' campaign (2014-2016), and Center for Disease Control's (CDC) ``Tips from Former Smokers'' campaign (2012) found success in using real stories and testimonials of former smokers with similar demographic characteristics as those of their target audience. Research suggests that this testimonial message strategy was effective in enhancing long-term and short-term disease susceptibility of tobacco use~\cite{c13}\cite{c14}\cite{c15}\cite{c16}. We also know that peer influence plays a critical role in altering health behaviors, especially among vulnerable populations such as adolescents~\cite{c17}.

In this paper, we design a framework based on social bots, named Notobot (short for No-Tobacco Bot), to address pro-tobacco expressions using a targeted testimonial communication strategy (Figure~\ref{fig:complete}). Notobot exposes a user identified as being pro-tobacco to a tweet posted by a real former smoker (referred to as message creators from hereon) who has the most similar user account characteristics (e.g. age of the account, temporal activity rates, social network structure, etc.). Our contributions are two-fold. First, we show that neural networks are more effective than classical methods (Logistic Regression, Decision Trees, and Support Vector Machines) for detecting tobacco proclivity, we believe in part due to the nuanced language concerning tobacco usage. For example, "I could use a smoke" shows a desire to do an action, not the ability to use an object. Second, we propose to counter pro-tobacco attitudes by leveraging machine-learning to crowd-source  anti-smoking messages from former smokers.  Notobot will, in the near future, undergo testing and evaluation among cohorts of participants who post pro-tobacco tweets. 

%% file: src/motivation.tex
\section{Motivation}
The goal of preventive medicine is to prevent the onset of unhealthy behaviors that contribute to overall morbidity and mortality. Public health intervention messages addressing unhealthy behaviors associated with chronic diseases such as lung cancer and diabetes, are often tasked with communicating immediate health risks of activities such as smoking or high sugar consumption. A recent tobacco education campaign, \textit{Tips From Former Smokers}, was successful in persuading youth about the immediate risks of smoking \cite{c13, c14, c15, c16}. This campaign profiled compelling stories of real smokers living with debilitating health effects from smoking and secondhand smoke exposure. Other campaigns found success in increasing awareness about diabetes by featuring stories of Latino youth advocates  and using youth-generated ``spoken-word'' messages in the campaigns~\cite{c26, c27}. A majority of national- or local-level health communication interventions are delivered either via community organizations (such as schools), mass media (television, radio), or paid advertisements on the digital media (Facebook, Twitter) that direct audiences to campaign websites. Examples of few notable exceptions that leveraged the unique advantages of social media include the Amyotrophic Lateral Sclerosis (ALS) "ice bucket challenge", and  \textit{Movember}'s campaign for men's health awareness~\cite{c28, c29}.

While there has been substantial focus on developing effective health messages, delivery of health prevention interventions calls for more reach and flexibility. Given the popularity of social media, smoking prevention efforts can continue to leverage the strengths of platforms such as Twitter. Social media platforms offer opportunities for cost-effective and scalable health interventions. These platforms also allow campaign strategists to track individuals' online expressions/trajectories and customize interventions by identifying the right time and content for delivery. 

%% file: src/method.tex
\section{Methodology}

\begin{figure}
  \includegraphics[width=\linewidth]{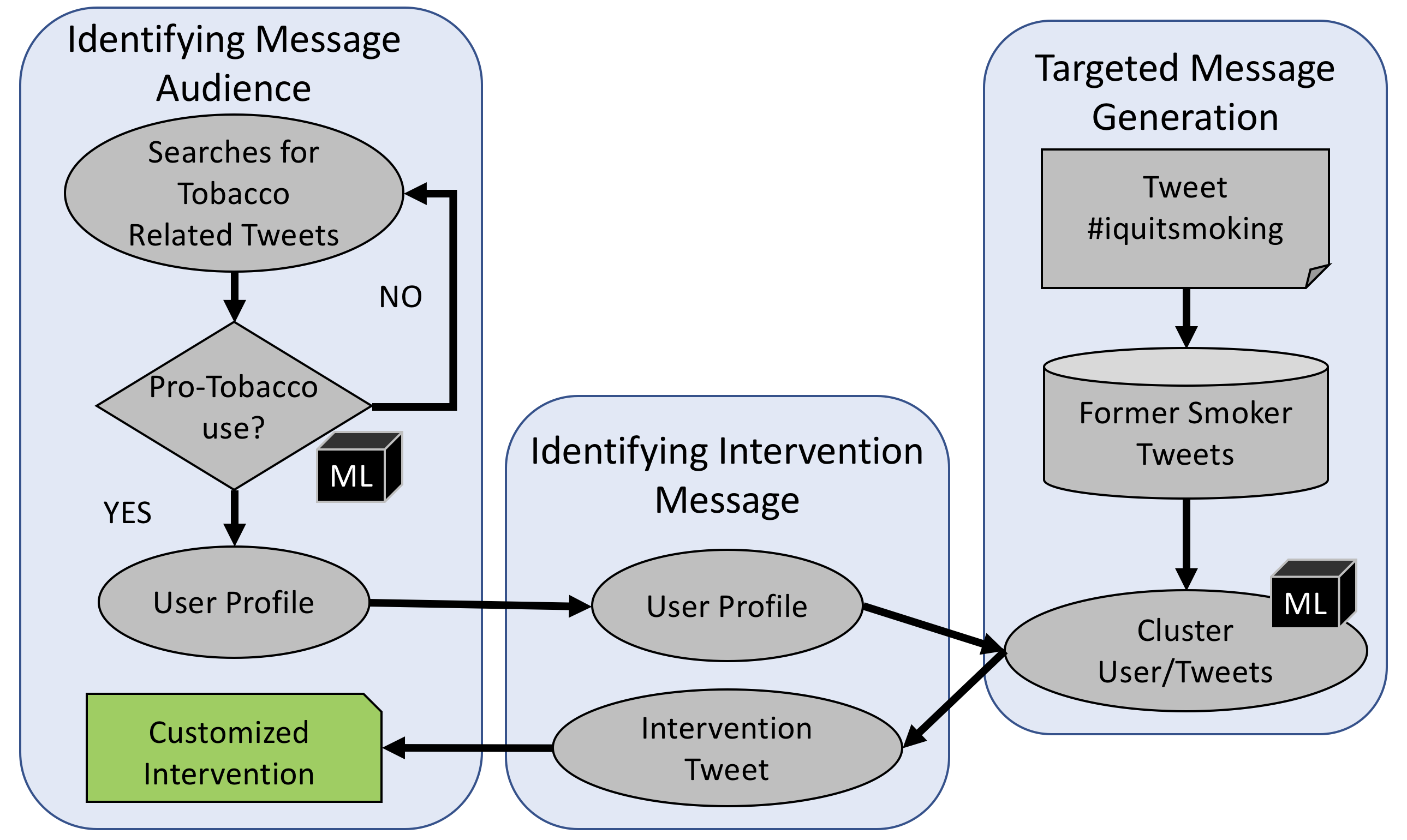}
  \caption{Overall Notobot system: A user who posts a pro-tobacco tweet will be exposed to a customized tweet about not smoking}
  \label{fig:complete}
\end{figure}

The first step in our framework is to identify the target audience in order to deliver an intervention message to the target. The target audience is identified in three steps: (1) Capture tweets with tobacco-related keywords in an automatated process, (2) Classify tobacco-related tweets into pro-tobacco and not pro-tobacco tweets using a data-driven model based on convolutional neural networks, (3) Extract user information associated with pro-tobacco tweets. Next, those users posting pro-tobacco tweets receive private Twitter message(s) consisting of real tweets posted by former smokers who have a similar profile as the pro-tobacco users. This is achieved in three steps: (1) Extract tweets and user information associated with hashtags about  smoking cessation (e.g., \#iquitsmoking), (2) Cluster user accounts in groups of similar users based on metadata, (3) Identify representative tweets associated with each cluster using unsupervised machine learning. Lastly, we expose the user who posted a pro-tobacco tweet with a customized intervention tweet sourced from former smokers who have similar user profile. The overall system architecture of Notobot is illustrated in Figure \ref{fig:complete}.


\subsection{Data Collection}
Data were obtained from the Twitter Streaming API using twitter-4J, an open source database used to collect and analyze data from the API. All tweets posted between January 1, 2018 to January 31, 2018 matching a set of tobacco-related keywords have been collected. These keywords, whose list is provided in the Appendix, appeared in the text of tweets or in an associated hashtag. The number of the initial samples consisting of tobacco-related tweets is 4,225. 


\subsection{Architecture }

\subsubsection{Extracting the metadata of message authors}
%
%
We obtained the message authors' metadata using Twitter's API. The metadata include: account creation date; the number of followers, followees (friends), lists, favorites, and posts; whether the account uses the default profile background, and default profile image; and whether it is a verified account. These metadata are  used next as features for clustering.

\subsubsection{Clustering message authors based on their metadata}
The clustering method consists of two phases---determining the number of clusters, and building a decision tree based clustering strategy. In the first step, we use k-means  to cluster the message authors based on each feature, and employ the silhouette score to get the best clustering solution. At the end of this step, after clustering, we obtain 14 different groups. If all features of two users were assigned to the same clusters, we labeled the two users as belonging to the same group. In the next step, we build a decision tree, based on Gini impurity scores, on the 14 groups discovered above. The decision tree  at each node is illustrated in Fig~\ref{fig:d-tree}. An intervention message is generated at each node.

\subsubsection{Extracting the metadata of the target user and classify the user based on the decision tree}
We identified targeted users using the tobacco-related keywords. Next, we extracted metadata associated with the targeted user profiles to classify them based on the decision tree created from the metadata of message creators. Finally, we selected a representative tweet from the message pool of the group as the intervention message for that specific group.

%
%
\begin{figure}
\centering

\subfigure[Intervention 1]{
\includegraphics[width=\linewidth]{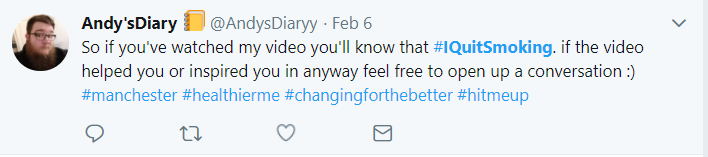}}
\subfigure[Intervention 2]{
\includegraphics[width=\linewidth]{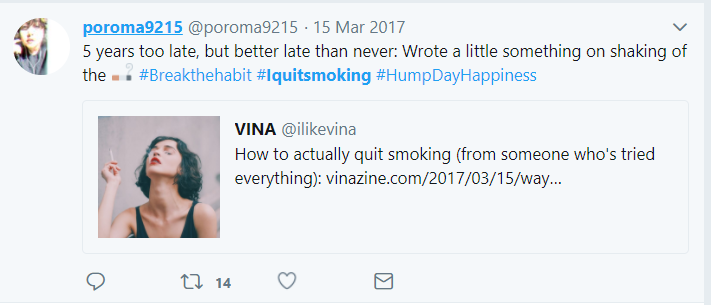}}
\caption{Example intervention tweets found from \#iquitsmoking}
\end{figure}


\subsection{Machine Learning Models}

\subsubsection{Logistic Regression}
Logistic regression is a regression model that is used for binary classification. It was developed by David Cox in 1958 \cite{c36} and has been used in a wide variety of applications and fields. We implement a binary model that predicts the odds of a tweet being pro-tobacco based on the embedding of the tweet. Given a generalized linear model:
\begin{equation}
h_{\theta}(x)=\frac{1}{1+e^-\theta^{T}x}
\end{equation}
The probability that a tweet is pro-tobacco or not pro-tobacco ($Pr(y|x;\theta)=1$, 100\% likely that is is pro-tobacco and $Pr(y|x;\theta)=0$, 0\% it is pro-tobacco) is modeled as: 
\begin{equation}
Pr(y|x;\theta)=h_{\theta}(x)^{y}(1-h_{\theta}(x))^{(1-y)}
\end{equation}
Which we then maximize the log likelihood using a gradient descent algorithm.
\begin{equation}
log L (\theta|x)=\sum_{i=1}^{N}log Pr(y_{i}|x_{i};\theta)
\end{equation}
Our implementation used the LogisticRegressor in the TensorFlow module \cite{c37}.

\subsubsection{Decision Tree}
Decision Trees are another classification method which are based on an underlying probability model \cite{c38}. Given the manually labeled data which consist of $y$, the label and $\vec{x}$, the embedding of the tweet, the data can be viewed as $(x_{1},x_{2},...x_{p},y)$ where each $x_{i}$ is a feature of the embedding. We applied the classification algorithm proposed in \cite{c39}. Working from the top and going down the tree, the Gini impurity which is a measure of how to best split the features is defined as:
\begin{equation}
I_{G}(p)=\sum_{i=1}^{K}p_{i}\sum_{i\ne j}p_{j} ; \sum_{i\ne j}p_{j}=1-p_{i}
\end{equation}

\subsubsection{Support Vector Machines (SVM)}

Support Vector Machines were first theorized in 1963 \cite{c35} and implementation algorithms used in this paper is from 1993 \cite{c34} . Given $n$ data points of the form $(\vec{x}_{1},y_{1})...(\vec{x}_{n},y_{n})$ where $\vec{x}_{i}$ is a $p$ dimensional vector that is the embedding of a tweet. Furthermore, $\vec{y}_{i}$ is either 1 for pro-tobacco tweet or -1 for non pro-tobacco tweet. 
SVM is a method to do a binary classification by developing a hyperplane $\vec{w}\cdot\vec{x}-b=0$ where $\vec{w}$ is the normal vector of the hyperplane and $b$ is a measure of the offset.  
In the non-linear implementation of soft-margin, the hinge loss function is shown in Equation 2 which is minimized using a sub-gradient descent algorithm. 
The $\lambda$ is a parameter that captures the trade off between margin-size and ensuring $\vec{x}_{i}$ in on the correct side of the margin. 
\begin{equation}
\left[\frac{1}{n} \sum_{i=1}^{n} max ( 0, 1-\vec{y}_{i}(\vec{w}\cdot\vec{x}_{i}-b))\right]+\lambda||\vec{w}||^{2}
\end{equation}

\subsubsection{Multi-Layer Perceptrons (MLP)}
MLP is a standard neural network approach in artificial intelligence that is trained using backpropagation. Other than the nodes in the input layer, the other nodes use a nonlinear activation function. Here we apply a simple MLP which only has one hidden layer \cite{c40}. In this analysis, we used the rectified linear unit (ReLU) activation function defined by \cite{c41}:
\begin{equation}
f(x)=x^{+}=max(0,x)
\end{equation} 

\subsubsection{Char-CNN}
Character-based convolutional neural networks (Char-CNN) treat text as a raw signal at the character level\cite{c20}. This leverages the temporal convolutional module:
\begin{equation}
h(y)=\sum_{x=1}^{k}f(x)\cdot g(y\cdot d-x+c)
\end{equation}
where $g(x)$ is a discrete input function and $f(x)$ is a discrete kernel function. The stride of the convolution $h(y)$ between $f(x)$ and $g(x)$ is $d$ and $c$ is an offset constant \cite{c20}. The Char-CNN was implemented using the TensorFlow module as well \cite{c37}.

\subsection{Identifying Message Audience}

In the identifying message audience (IMA) section of Notobot, we aim to find the target audience for the intervention, i.e. users who author pro-tobacco tweets. These users are considered as a set of targets we want to find and intervene to modify their smoking related attitudes (e.g. increase perceptions of risk and decrease perceptions of benefits from smoking), ultimately changing smoking behaviors (e.g. smoking cessation). In IMA, we first searched for tobacco-related tweets based on tobacco-related keywords. Each tweet with tobacco-related keywords was fed into a pre-trained classifier to decide whether it was pro-tobacco or not. If Notobot did not classify the tweet as pro-tobacco, the tweet was discarded and the next tweet was analyzed. If the tweet was pro-tobacco, we retrieved the corresponding user profile using Twitter API and sent it to identifying intervention message module (IIM) for the customized intervention.

To train the pro-tobacco classifier, we collected a number of tobacco-related tweets and manually labeled (pro-tobacco or not pro-tobacco) each tweet. Next, we considered different models for classification. We determined that although many existing models including logistic regression, decision tree, support vector machine (SVM), and multi-layer perceptrons (MLP) have been studied for the binary classification, we implement character-based convolutional neural networks (Char-CNN)~\cite{c20}. Char-CNN offers several advantages. First, it is more robust to handle out-of-dictionary words. When natural language texts are handled, tf-idf, bag-of-words, and word embedding model~\cite{c30} are well-known methods to represent the texts into numerical features. These are word-based methods, which are not robust to handle typos, slang, abbreviated words, infrequent words, and buzzwords which are especially widely used in tweets. For example, word embedding model is only able to deal with pre-seen words and tend to ignore words that have not been used for training the embedding model. Second, the convolutional filters are appropriate to extract localized features. This characteristic is especially desired for texts in tweets. For example, if someone uses ``grrrreat'' instaed of ``great'', the convolutional filter can capture the similarity between the words. Since many words are written informally in tweets, Char-CNN is more suitable. Table~\ref{tab:acc} provides the classification accuracy on 4,225 tweets with labels (70\%/30\% split for training and testing sets) from different classifiers.

First, we quantized a set of characters by prescribing an alphabet of size $m$ into one-hot encoding. Once a lookup layer was used to transform the one-hot encoding to fixed length embeddings. Following alphabets including a blank space are what we used:

\begin{center}
\texttt{abcdefghijklmnopqrstuvwxyz0123456789}\\
  \texttt{'"`-,;.!?:$\backslash$/|\_@\#\$\%\^\*$\sim$+=<>()[]\{\}}
\end{center}
Any characters that are not in the pre-defined alphabets were quantized as all-zero vectors.

\begin{align}
x^{(l+1)} = W*x^{(l)}
\end{align}
where * is a convolution operator and $x^{(l)}\in\mathbb{R}^3$


\subsection{Targeted Message Generation}
In the target message generation (TMG) section of Notobot , we generate the best matched intervention messages for the target audience. We collected anti-tobacco tweets posted by former smokers, and treated those tweets as our initial pool of intervention messages. Once we identified a target Twitter user posting pro-tobacco tweets, TMG selected an intervention message from a pool based on a matching mechanism. 

To identify message creators, we searched through Twitter using tag $\#iquitsmoking$, under which tag many people shared their experiences about quitting smoking. We collected 216 users' tweets with this tag as the initial pool for our matching mechanism. The matching mechanism involved the following procedure: (i). extracting metadata of message creators. (ii). clustering the message creators based on their metadata. (iii). matching the target audience with a post from a message creator pulled from, or identified with, the cluster.

%% file: src/results.tex
\section{Results}

\subsection{Identifying Message Audience}
Once Notobot identifies a tweet with specified keywords, the system needs to determine if it is pro-tobacco use or not. Table \ref{tab:acc} shows the results of various classifiers on the 4,225 manually labeled tweets. The Char-CNN was the best performing model used for implementing Notobot. In the future, tweets determined to be pro-tobacco will be manually verified and used to update the classifier. 

\begin{table}[!t]
\centering
\caption{Different Classifiers of Pro-Tobacco Tweets
\label{tab:acc}} 
\begin{tabular}{lcc}
\hline%
\textbf{Classifier}&\textbf{Accuracy}&\textbf{Standard Deviation}\\%
\hline%
Logistic Regression &	0.6767	&	0.0807	\\
\hline
Decision Tree	& 0.6436	&	0.0554	\\
\hline
SVM	&	0.6904	&	0.068	\\
\hline
MLP	&	0.6968	&	0.0674	\\
\hline
Char-CNN	&	0.7401	&	0.064	\\
\hline%
\end{tabular}%
\end{table}

\subsection{Target Message Generation}
\begin{figure}
  \includegraphics[width=\linewidth]{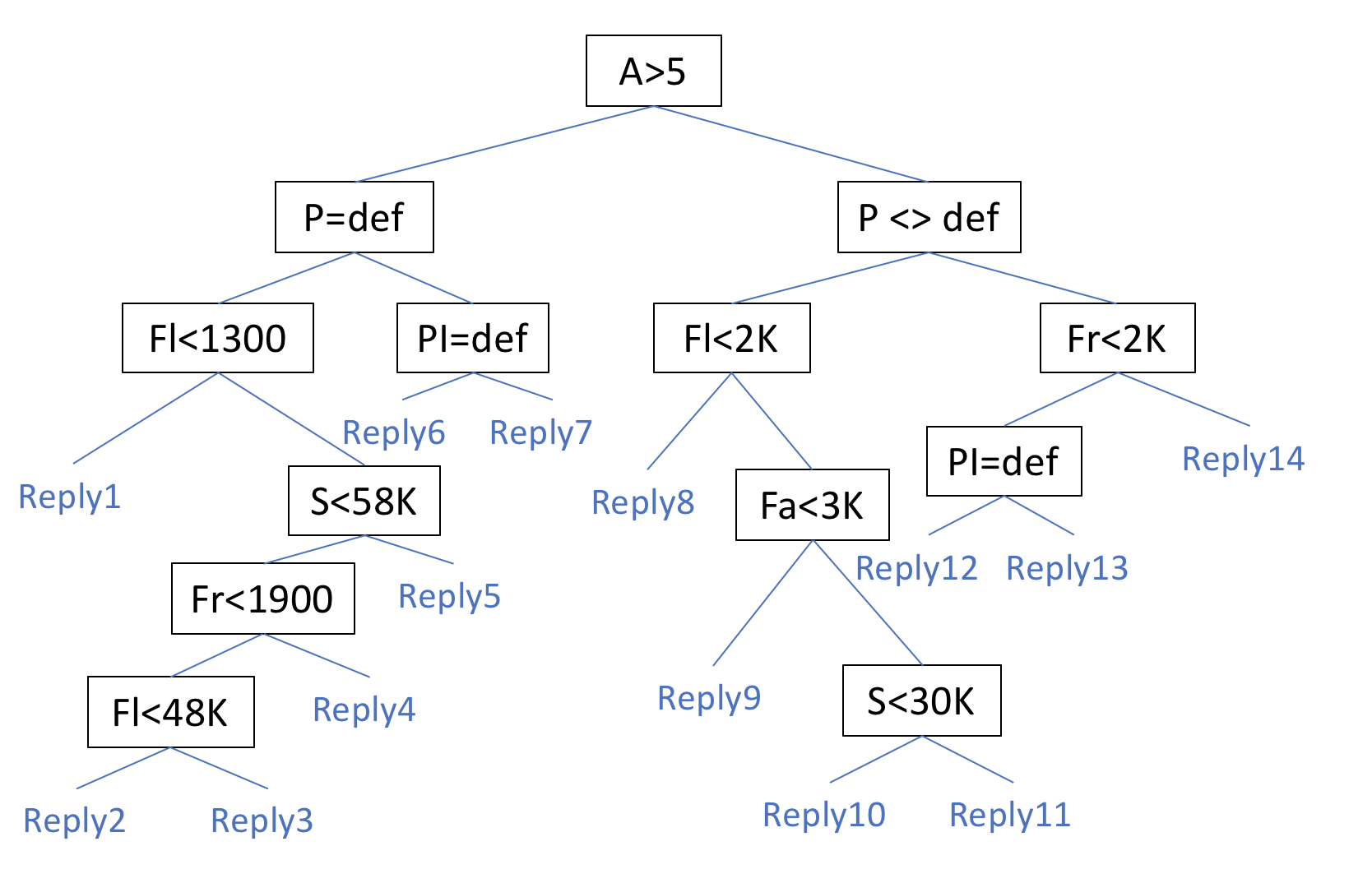}
  \caption{Decision Tree showing profile features as splitting factors for classifying users into different groups; the left branch is always TRUE and the right branch is always FALSE. The abbreviations at each node correspond to the features reported in Table \ref{tab:user}.}
  \label{fig:d-tree}
\end{figure}

%
%
From the 216 \#iquitsmoking tweets we identified 9 dimensions of those user’s metadata from which to cluster their tweets on. The dimensions, their data type and justification are listed in Table \ref{tab:user} These dimensions represent the returned user metadata from the search API which provide useful insight. By doing a series of independent clusters, we constructed a decision tree that maps user profiles to an Intervention bin. There are 11 bins which are highlighted in Figure \ref{fig:d-tree} determined by user dimensions from \ref{tab:user} variables in parenthesis.  When remapping the training set, all samples would map to the bin containing the tweet that they had posted. Tailored interventions have been shown to be superior to one-size fits all interventions \cite{c44}. As a result, the present intervention utilized a post describing the benefits of quitting and/or how to quit smoking from a former smoker with similar profile characteristics to the user receiving the intervention.  

\begin{table}[!t]
\centering
\caption{Different Classifiers of Pro-Tobacco Tweets
\label{tab:user}} 
\begin{tabular}{llc}
\hline%
\textbf{User Aspect}&\textbf{Definition}&\textbf{User Dimension}\\%
\hline%
created\_at\_mms 	&	Age of account in months	&	(A)	\\
favourites\_count 	&	Breath of activity	&	(Fa)	\\
followers\_count 	&	More of a social leader	&	(Fl)	\\
friends\_count 	&	Social network size	&	(Fr)	\\
listed\_count 	&	Popularity	&	(Lc)	\\
statuses\_count 	&	Level of activity	&	(S)	\\
default\_profile 	&	Advanced user	&	(P)	\\
default\_profile\_image 	&	Advanced or transparent user	&	(PI)	\\
verified 	&	Very active user	&	(V)	\\
\hline%
\end{tabular}%
\end{table}

\subsection{Notobot Demonstration}

\begin{figure}
\centering
  \includegraphics[width=\linewidth]{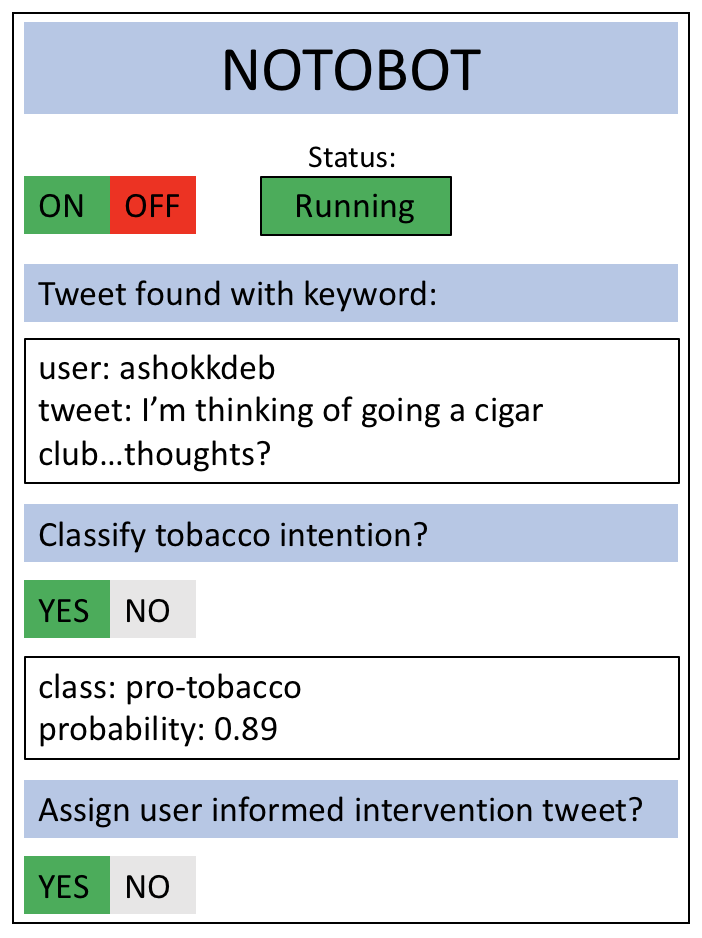}
  \caption{User Interface of Notobot}
  \label{fig:Notobot}
\end{figure}

The image in Figure \ref{fig:Notobot} shows the user interface of the system. On clicking the main Notobot, the system shows that it is in an on-state and is actively searching Twitter for new tweets with tobacco-related keywords. Due to current API restrictions, Notobot does a new scan every minute. Once a new tweet is identified, it is presented to the user with the option of classifying it as pro-tobacco or not. If yes, then the Char-CNN classifier from above will determine if it pro-tobacco or not. If it is pro-tobacco, the user is presented with the option of determining a customized intervention. If yes, the decision tree from before will collect the user profile of who posted the tweet and then map it to a bin of Interventions using the user metadata. One of those AI selected tweets will be presented with the option of posting to the user. Currently, Notobot only posts to its own wall as the university Institutional Review Board (IRB) process was not needed for Notobot's development. We are in the process of working with the IRB to deploy and test Notobot in a controlled experiment. The step-by-step process presented here is for demonstration purposes only and has the ability to operate in a fully automated mode.

The current implementation of Notobot demo website is hosted on Apache server. On the Notobot web page, there are radio buttons, which have been enhanced with toggling capabilities for the following operations.
\begin{enumerate}
\item  Switch On/Off Notobot python script for searching  possible pro-tobacco tweets.
\item To run the classifier python script to identify whether the tweet is pro-tobacco or not.
\item To run the customized intervention python script and post the intervention message on Twitter.
\end{enumerate}

The event handling part on the different buttons is done via javascript and php. The function "exec()" in php allows us to run any script the way we do using a shell. And the output of the function can be saved in a php variable which can used for display purpose or can be passed as a parameter to some other php function.

\begin{figure}
\centering
  \includegraphics[width=\linewidth]{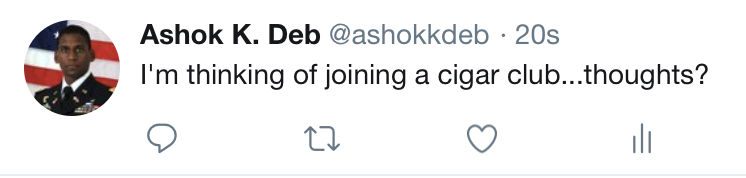}
  \caption{Example of Tobacco related tweet found by Notobot}
  \label{fig:tob1}
\end{figure}

\begin{figure}
\centering
  \includegraphics[width=\linewidth]{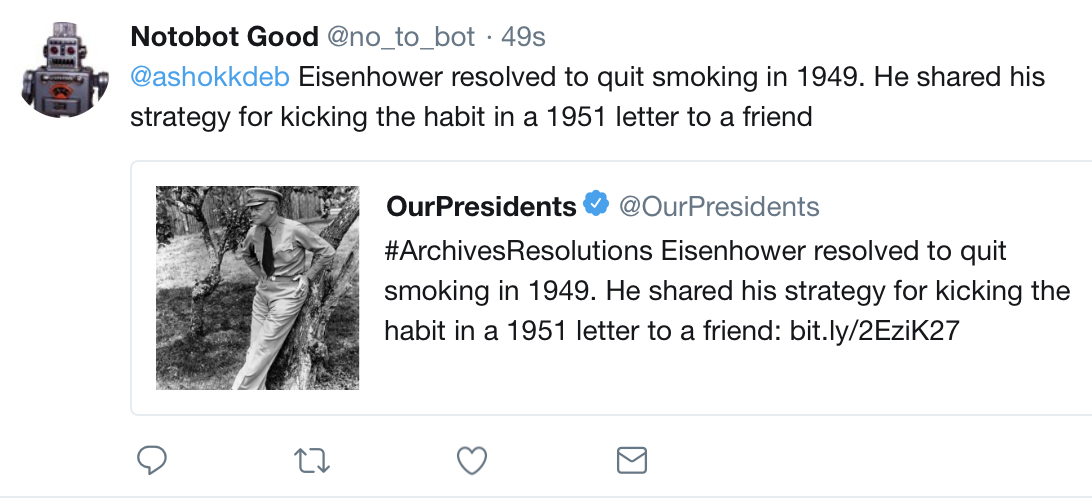}
  \caption{Example of Customized response by Notobot}
  \label{fig:tob2}
\end{figure}

Figure \ref{fig:tob1} shows an example of a tweet that contains a tobacco key word that Notobot found. Upon analyzing the tweet it was found to be a pro-tobacco tweet with 95\% confidence. Using the user profile information of the user that posted the tweet, the AI recommended tweet is from Intervention Group 7. In Figure \ref{fig:tob2} you can see that Notobot posted the recommended tweet while mentioning the user. This test case was used for demonstration purposes and used the account of one of the authors. No interaction took place with other Twitter users. That is left for future work in an intervention study after an Institutional Review Board has been conducted.

%% file: src/related.tex
\section{Related Work}
\subsection{Machine Learning in Behavior Classification}

Previous work shows content and sentiment analysis of tobacco-related Twitter posts and builds machine learning classifiers to detect tobacco-relevant posts and sentiment towards tobacco \cite{c18}. Additionally, the limitations of the machine learning classifier have been explicitly mentioned by the authors as their future work; however, it does not talk about any concrete intervention. \cite{c19} proposes a supervised approach to identifying sentiments and information dissemination concerning e-cigarettes from Twitter data. Again, it is limited to analysis and doesn't provide solution for scalable public health interventions. 

\subsection{Social Media as an Intervention Platform}
Smoking prevention through mass media \cite{c25} was heavily studied before the Internet and social media became widespread. Since that time, social media has been used a venue for health communications across a number of issues such as mental health \cite{c22} , HIV prevention \cite{c23}, and increasing physical activity \cite{c24}.

While slightly dated, the analysis from Chou et. al. \cite{c21} is certain to be true today. Both online presence and online use is not uniformly distributed across age groups and other demographics. This is important since the age group of the initial target group for intervention includes adolescence (age 13-18) and young adults (age 18-24). This group focuses on the most active age demographics on social media. While Twitter may not be the best social media platform for this age group, Notobot can still serve as a test case whose algorithm and design can be ported to other social media platforms. \cite{c21} also highlights how our work is designed only for the US population and only for those who tweet in English as the reach for other communities would be a different distribution.

%% file: src/keywords.tex
\begin{table}[!ht]
\centering
\caption{Keywords used to search for tobacco-related tweets
\label{tab:keywords}} 
\begin{tabular}{lll}
\hline%

ecig	&	secondhand vape	&	copenhagen	\\
e-cigs	&	secondhand vaping	&	camel	\\
ecigs	&	second-hand vape	&	snus	\\
e-cigarette	&	second-hand vaping	&	pall mall	\\
ecigarette	&	vape smoke	&	newport	\\
e-cigarettes	&	ecig smoke	&	wakeup	\\
ecigarettes	&	e-cig smoke	&	cheerupbigtobacco	\\
vape	&	e-cigarette smoke	&	transformtobacco	\\
vaper	&	vape shs	&	swishersweets	\\
vaping	&	ecig shs	&	swisherartistproject	\\
vapes	&	vape secondhand smoke	&	swisherartistgrant	\\
vapers	&	vape second-hand smoke	&	swisheratl	\\
nicotine	&	esmoke	&	blunation	\\
tobacco	&	e-smoke	&	blucigs	\\
cigarette	&	stillblowingsmoke	&	justyouandblu	\\
cigarettes	&	still blowing smoke	&	plusworks	\\
cigar	&	notblowingsmoke	&	thisfreelife	\\
atomizer	&	not blowing smoke	&	swishersweeties	\\
atomizers	&	capublichealth	&	swishermusiccity	\\
cartomizer	&	tobaccofreekids	&	tobacco21	\\
cartomizers	&	notareplacement	&	sbx27	\\
ehookah	&	trulyfree	&	sbx25	\\
e-hookah	&	truly free	&	FDAdeeming	\\
ejuice	&	sb140	&	FDAtobacco	\\
ejuices	&	sb 140	&	prop56	\\
e-juice	&	sb24	&	prop64	\\
e-juices	&	sb 24	&	juul	\\
eliquid	&	cherry tip cigarillos	&	smokeless	\\
eliquids	&	mini-cigarillos	&	ryo	\\
e-liquid	&	tip cigarillos	&	rollyourown	\\
e-liquids	&	king edward cigars	&	heatnotburn	\\
blu	&	royal gold cigars	&	glo	\\
njoy	&	sweet coronella	&	iqos	\\
green smoke	&	swisher blk	&	HnB	\\
south beach smoke	&	swisher sweets	&	reduced risk products	\\
eversmoke	&	vapercon	&	british american tobaccco	\\
joye 510	&	vapercon west	&	PMIScience	\\
joye510	&	grimmgreen	&	InsidePMI	\\
joyetech	&	vapor	&	PMI	\\
lavatube	&	electronic cigarette	&	BTI	\\
lavatubes	&	vape meet	&	switchfromsmoking	\\
logicecig	&	EcigsSaveLive	&	smokethistoo	\\
logicecigs	&	EcigsSaveLives	&	cigarettesaredead	\\
smartsmoker	&	EcigsSavesLives	&	nomorecigarettes	\\
smokestiks	&	vapecon	&	vapepromote	\\
v2 cig	&	fresh empire	&	stopsmoking	\\
v2 cigs	&	freshempire	&	switchtovaping	\\
v2cigs	&	camel crush bold	&	vapingkicksash	\\
v2cig	&	camelcrushbold	&	papastratos	\\
mistic	&	menthol	&	vapenews	\\
21st century smoke	&	clove	&	ploomtech	\\
logic black label	&	hookah	&	heets	\\
finiti	&	cigarillo	&	failbycigs	\\
nicotek	&	blunt	&	Phix	\\
cigirex	&	vaporcade	&	7000chemicals	\\
logic platinum	&	narguile	&	therealcost	\\
cigalectric	&	shisha	&	tobaccofreelife	\\
xhale o2	&	marlboro	&	freshevents	\\
cig2o	&	vuse	&	everytrycounts	\\
green smart living	&	swisher	&		\\
krave	&	black and mild	&		\\

\hline%
\end{tabular}%
\end{table}